\documentclass{emulateapj}

\shorttitle{MAPPING INSPIRAL RATES ON MODEL PARAMETERS }
\shortauthors{O'Shaughnessy, Kalogera, \& Belczynski}

\begin{document}

\title{Mapping  inspiral rates on population synthesis parameters} 
\author{R.\ O'Shaughnessy, V.\ Kalogera, \& K.\ Belczynski}

\affil{Northwestern University, Department of Physics and Astronomy,
  2145 Sheridan Road, Evanston, IL 60208, USA}
\email{oshaughn, vicky, belczynski@northwestern.edu}

\begin{abstract}
Formation rates of compact-object binaries are often derived from population synthesis calculations.  However, such calculations  depend sensitively on a relatively large number of model input parameters.  Given considerable uncertainty in those model parameters,   the predicted inspiral rates for double compact objects  relevant to gravitational-wave interferometric detectors have been shown to be are uncertain by several orders of magnitude.
Typically, inspiral rates are estimated for only a small set of models with a remarkably poor coverage of the highly multi-dimensional parameter space (primarily because of limited computer resources). 
Here, using as an example  seven population-synthesis model parameters, we show that it is possible to derive fits of double-compact-object inspiral rates dependent simultaneously on all seven parameters. We find these fits to be accurate to 50\% for binary black holes and to 40\% for binary neutron stars.
The availability of such fits implies that (i) depending on the problem of interest, it is not necessary to complete large numbers of computationally demanding population synthesis calculations; and (ii) for the first time, the sufficient exploration of the relevant phase space and the assessment of the uncertainties involved is not limited by computational resources and becomes feasible.
\end{abstract}


\section{Introduction}

Over the last decade the question of the Galactic inspiral rate of binaries with two compact
objects (neutron stars NS or black holes BH) has attracted attention primarily because of the
development and planning of gravitational-wave interferometric detectors both on the Earth and
in space (e.g., LIGO, VIRGO, GEO600 and LISA). Such rate estimates have been widely used in the
assessment of gravitational inspiral detectability, given assumed instrument
sensitivities \citep{CutlerThorne2002}.

A number of different groups have calculated inspiral rates using population synthesis
calculations, most commonly with Monte Carlo methods  
\citep{Fryer,PortZwart,BetheBrown,FWH,
StarTrack,VT}. Such studies
consider the complete formation history of double compact objects through long sequences of
binary evolution phases. However, our current understanding of single and binary star evolution is
incomplete.   Therefore uncertainties are often parameterized and some of these parameterizations
are better constrained (usually empirically) than others.  Depending on the binary type of
interest, rate predictions are sensitive to different parameters and often degeneracies have
been shown to exist. 
It is clear that a comprehensive a priori estimate of inspiral rates
requires estimates of model uncertainties as well. However, current binary population synthesis
codes are computationally demanding (depending  on their level of sophistication) and
covering a large part of the multi-dimensional parameter space has been proven  very challenging. As a
result, earlier studies have estimated model
uncertainties by varying only one or two parameters at a time, and considered at best a couple
of dozen of models.

%
In this {\em Letter} we discuss ways to reduce the computational cost of population synthesis
calculations (used here to focus on the formation rates of
double compact objects).  As a result, significantly more extended parameter-space
searches become feasible.   We make use of  the basic concept of {\em genetic algorithms}, which are ideal for the exploration of
highly multi-dimensional spaces.   We validate the suggested ways and use them to examine whether
the derivation of fits for the inspiral rates as a function of a large number of free
parameters is possible. We find that, while strong correlations between model parameters are
apparent, it is possible to fit inspiral rates to an acceptable accuracy
(50\% in this study, but possibly smaller if the original simulations used to derive the fits are more accurate). 
 In what follows we first describe briefly the population synthesis
code we use in our analysis. We then discuss the basic concepts of genetic algorithms and
describe how they can be used to increase the computational efficiency of synthesis codes. We
conclude with the derivation of fits for NS-NS and BH-BH rates and discuss the implications of
the availability of such fits for deriving constraints on BH-BH rates in the future.

\section{\label{sec:code}StarTrack population synthesis code}
%
%

To generate and evolve stellar populations until double compact-object formation occurs, we use
the \emph{StarTrack} code first developed by Belczynski, Kalogera, and Bulik
(2002) 
[hereafter BKB] and recently significantly updated and tested as described in detail in
Belczynski et al.\ 2004.

%
%

Here we briefly summarize the main parts of the computations relevant to this study. We generate a large
number $N$ of binaries specified without loss of generality by the mass $m_1$ of the primary (in
solar masses $M_\odot$);
the mass ratio $q=m_2/m_1\le 1$ between the primary and secondary objects;
the semimajor axis $A$ (in solar radii);
 and the orbital eccentricity $e$.  
We draw  binaries from the distributions given in BKB Eqs.\ (2), (4), (5), and
\begin{equation}
\rho(q) dq \propto  
 \cases{1, &  $q\in[0,q_c]$ ; \cr
       (q/q_c)^{-r}, &  $q\in[q_c,1]$   .\cr
}
\label{eq:rhoq}
\end{equation}
where $q_c=0.2$ and $r\in[0,3]$ is a model parameter; for a discussion see Kalogera \& Webbink (1998).\footnote{To improve our computational efficiency, we restrict the Monte-Carlo generations to  $m_1 > 4$ and $m_1 q > 4$, since we are interested in just NS and BH in our present study.}
We adopt a uniform star formation rate appropriate for our Galaxy \citep{SFR,SFR2}: each binary is assigned a
birth time from a uniform distribution between the formation of our galaxy ($t=0$) and the
present (assumed at $t=T=10$Gyr).  We evolve each binary until either it ceases to be a binary, or
until the present, whichever comes first.

During this evolution of $N$ massive binaries, some number $n$ of a certain type of events
occurs, corresponding to a rate of $r=n/T$ in the simulation.  To
obtain the time-averaged event rate ${\cal R}=T^{-1}\int (dn/dt) dt$ for the Galaxy, we scale
this computational rate up by a 
scale factor $s$
\begin{equation}
{\cal R} = s \times n/T
\end{equation} 
where $s$ is a ratio of the number of stellar systems we have
effectively simulated to the number of stellar systems  of the same type in the target
system (here, the Milky Way).   The appendix describes in
detail how we determine this absolute normalization scale factor.

%
%

\emph{Parameter Phase Space}: For the study presented here we have chosen to vary seven (7)
model parameters in the synthesis calculations. The choice is strongly guided by our past
experience with double-compact-object population synthesis (BKB) and represent the model parameters for which strong dependence has been confirmed. The list of these seven
parameters is as follows (for more details see BKB): the fraction of transferred mass that is lost from the binary in phases of
non-conservative mass transfer $f_a\in[0,1]$;
the common envelope efficiency coupled with the uncertainty in the donor star central
concentration $\alpha\times\lambda\in[0,1]$; three parameters describing the locations and
relative weight of two Mawellians in a bimodal NS kick distribution ($v_1\in[0,200]$km/s,
$v_2\in[200, 1000]$km/s, $s\in[0,1]$); a parameter describing
the stellar wind strength $w\in[0,1]$ relative to the reference-model assumptions; and the
power-law exponent $r\in[0,3]$ in the probability
distribution for the mass ratio $q=m_2/m_1$ [Eq.~(\ref{eq:rhoq})].
%
%
The remaining parameters are fixed and as described for our reference model in BKB.  Most notably, we use a binary fraction $f_b= 1$ (i.e. all stars are binaries) and solar metallicity ($Z=0.02$).

%
%

\emph{Synthesis Runs to Chosen Relative Accuracy}: Since we explore an
 exceptionally large database of models  to derive a fit for
the inspiral rate, we need to minimize the considerable
computational cost associated with each model. We  do so by
fixing the relative  statistical accuracy of the inspiral event
rate calculation from each model.

 Since the inspiral rates are linearly proportional to the
number of relevant events $n$ occurring in a given run, the relative accuracy of the inferred
rates is proportional to $\sqrt{n}$. For example, to obtain rates accurate within $\sim30$\%,
we require $\simeq 10$. Therefore, for every model, we
evolve each binary in succession, and
we stop generating new binaries when either (i) a \emph{fixed} number
$n$ of events has occurred or (ii) the total number $N$ of generated binaries grows above a
chosen threshold $N_{\max{}{}}$. Our choices for the two specific thresholds, $n$ and
$N_{\max{}{}}$, vary depending on the goal and nature of our calculation and are described in
what follows.

\subsection{\label{sec:nsSamp}Example: Sampling the NS-NS Rate}
As an example of the above process, we choose random combinations of
our $7$ population synthesis model parameters; for each combination,
we select binaries according to the chosen initial
distributions; we evolve each binary in succession and \emph{record}
the result at the end of the evolutionary sequence; we stop when we
reach either $n=10$  NS-NS binaries which merge within the simulation time $T=10^{10}$ yrs\footnote{For
brevity, we denote systems which merge due to gravitational radiation before the end of
the simulated interval by ``merging binaries''.} or when we have sampled $N=10^5$ binaries,
whichever comes first. Over the course of a month, on $\sim 10-15$
dedicated (of currently top-level speed) CPUs, we were able to
evaluate 488 different population synthesis models.

Each of these 488 runs provides a single,
relatively low-accuracy (30\%) but reasonable estimate of the NS-NS merger rate appropriate to the assumptions used in
that run. This collection  provides an unbiased sample of the NS-NS merger rate as a function
of population synthesis model parameters in a 7-dimensional space.  
 In Sec.~\ref{sec:fits}, we use these runs (along with other data) to generate \emph{fits},
  with which we can crudely and quickly estimate the NS-NS merger rate as a
function of model parameters.





\section{\label{sec:part}Accelerating Population Synthesis Simulations}

While we successfully estimate the NS-NS merger rate with the direct approach outlined above
for a large enough number of models,
the case of the BH-BH merger rate is significantly more challenging because they are typically
smaller than the NS-NS merger rates [cf.~Fig.~(\ref{fig:Histogram})].
Conversely this implies that with the progenitor-generation scheme in the synthesis simulations
the initial binary population is {\em greatly} dominated by systems which
\emph{do not} evolve to BH-BH binaries  (and especially \emph{merging} BH-BH binaries). 

We propose here that it is actually possible to use the above characteristics to our benefit
and derive appropriate {\em partitions} (i.e., constraints) that eliminate the majority of
irrelevant progenitors, but at the same time do not eliminate a significant fraction of
relevant progenitor. Although we concentrate here on the case of merging BH-BH binaries, it is
clear that the method we describe below can be applied to any type of compact object binaries
with low formation rates (relative to other binaries).

\subsection{Partitions on the Initial Binary Parameter Space for Merging BH-BH Formation}

To accelerate the synthesis simulations for BH-BH merger rates we look  generally for \emph{partitions} in the parent parameter space: surfaces in the  parameter space of all possible progenitors characterized by $P=(m_1, q, A,e)$ that separate interesting from uninteresting initial binaries.  

To derive a partition well-suited towards
finding progenitors of merging  BH-BH binaries,  we search for some   combination of parameters
$C=(c_1,c_2, c_3, c_4,c_5)$ such that the function
\begin{eqnarray}
p(P,C)&\equiv &c_1 \log_{10} (m_1) +c_2 \log_{10} (m_2) \nonumber \\
 &&  + c_3 \log_{10}  (A) + c_4 e - c_5
\label{eq:pt}
\end{eqnarray}
is both (i) positive ($p>0$) for \emph{all} BH-BH binaries
 and (ii) negative ($p<0$) for as many other non-BH-BH binaries as possible.
 [The form we use in Eq. (\ref{eq:pt}) for $p$ is an arbitrary choice, having no motivation
 besides convenience.]
Assuming such a partition can be found, we can safely ignore any progenitor parameters for which
 $p<0$; doing so speeds up each run by a factor equal to the ratio of ``weighted volumes''
 between the regions with $p>0$ and with $p<0$.\footnote{Here the volume is weighted by the
 assumed probability distributions of the initial parameters, rather than geometric coordinate volume.} Clearly this
 process is beneficial only if the speed-up factor defined here is significantly higher than
 unity.

Conceptually, these partitions can be found by a straightforward calculation.  For example, we can
take the recorded data from our 488 simulations and split the records into three groups: $A$ (the
set of all progenitor binaries which end up as merging BH-BH binaries), $B$ (the set of all binaries
which end up as merging NS-NS binaries), and $E$ (everything else).  Then we use some search algorithm to
find those parameters $C$ which have $p>0$ for as many members of $A$ as possible and,
furthermore, keep $p<0$ on a significant fraction of all members of $B$.

To be quantitative, we use a robust, \emph{genetic-algorithm} based search to find
combinations $C$ which maximize the following quantity:
\begin{eqnarray}
\label{eq:ptmax}
S(C)&\equiv& \frac{100}{N_A} \sum_{C\in A} \theta(p(P,C)) +  \frac{1}{N_B}\sum_{C\in B}
\theta(-p(P,C)) \\
 &=& 101 - \frac{100}{N_A} \sum_{C\in A} \theta(-p(P,C)) -  \frac{1}{N_B}\sum_{C\in B}
\theta(p(P,C)) \nonumber
\end{eqnarray}
where $\theta(x)$ is a step function: $\theta(x)=1$ if $x>0$ and 0 otherwise; and $N_A$ and $N_B$ are the numbers of
binaries in $A$ and $B$, respectively.  
For example, the second term $\sum_{C\in A} \theta(-p(P,C))/N_A$ represents the fraction of
elements of $A$ which are misclassified by $p$.
The ratio of the two prefactors, 100:1, is chosen so that a $1\%$ error in classifying
  binaries of type A would count as significant as a $100\%$ error in classifying binaries of
  type B.

\subsection{Requirements and Consistency tests}

The surprising fact is not that a partition like the one described above exists (i.e., not that
a maximum of Eq. (\ref{eq:ptmax}) can be found), but 
that it turns out to be both (i) \emph{useful}: the ratio of the weighted volumes with $p<0$ and
$p>0$ is surprisingly large,  about 9:1,
and so is the speed-up factor; and (ii) 
\emph{accurate}: the fraction of binaries with $P\in A$ that are misclassified and their evolution
is mistakenly ignored (i.e., $p<0$) is very small (for our case, it is only $0.14$\%!).

Of these two factors, the high accuracy is the most surprising. The partition we derive remains
accurate \emph{for each model examined}: we misclassify and ignore (i.e. $p<0$) no more than a 
small fraction  of the merging BH-BH binaries in \emph{any given run}.  Quantitatively, if we consider any
simulation $k$ in our set of 488 runs and denote by $A_k$ the set of all BH-BH binaries in that
run, then $\sum_{C\in A_k} \theta(-p) < 0.12 N_{A_k}$; further, only $5\%$ of runs have
  errors $>5\%$.

\subsection{Application: BH-BH runs via partitions}

As described above, we found a surprisingly robust partition which efficiently and accurately
rejects the progenitors of binaries which do not evolve into merging BH-BH binaries
[cf.~Eq.~(\ref{eq:pt}), using the coefficients given in the first line of Table~\ref{tab:partitions}].  Thus, to
reduce the computational burden needed to perform population synthesis runs geared towards the formation merging BH-BH binaries, we use
this partition to augment the general procedure outlined in Sec.~\ref{sec:code}: we select
progenitor parameters according to the BKB distributions; we reject some progenitors on the
basis of this partition; we evolve each progenitor parameter combination in succession,
stopping when we reach either $n=10$ merging BH-BH binaries or $N=5\times 10^6$ total
binaries sampled, whichever came first. 
 The second line in Table~\ref{tab:main} summarizes these choices.

As in the NS-NS case, we chose random combinations of our 7 population synthesis model
parameters.  We rather quickly were able to evaluate 312 such
combinations.

\begin{deluxetable}{cccrr}
\tablecolumns{5}
\tablecaption{Population Synthesis Runs}
\tablehead{ \colhead{Target Event} &\colhead{Attempted} & \colhead{Complete} & \colhead{n} &
\colhead{$N_{\max{}{}}$} } \startdata {\rm NS-NS(a)} &488 & 418 & 10 & $10^5$ \\ {\rm BH-BH}
&312 & 273 & 10 & $5\times 10^6$ \\ {\rm NS-NS(b)} &151 & 123 & 100 & $5\times 10^6$%
\enddata
\label{tab:main}
\end{deluxetable}






\subsection{\label{sec:nsns}Application: NS-NS runs via partitions}
As with BH-BH mergers, we can similarly attempt to employ partitions to reject systems which
cannot possibly evolve into merging NS-NS binaries.  Unfortunately, in sharp contrast to the
BH-BH case, for NS-NS mergers we have not been able to 
dramatically accelerate each run -- even with the use of
several partitions simultaneously (i.e., we reject a system if  \emph{any} partition is
negative), each designed to filter out a specific
type of contaminant (e.g., WD-WD binaries, 
disrupted binaries, WD-NS binaries, BH-BH binaries, etc; see~Table~\ref{tab:partitions} for the specific partitions used).  Nonetheless, the identified partitions remained
quite accurate (average error probability $1.3$\%) and offered a non-negligible improvement in
computation speed: a speed factor of 2.5, based on the ratio of ``weighted'' volumes], we performed an additional set of 151 runs,
once again randomly distributed over model parameter space,
designed to provide higher-accuracy ($n=100$, leading to a statistical accuracy of 10\%) estimates of the NS merger rate.  
The third line in Table~\ref{tab:main} summarizes these choices.

\section{Results: Merger Rates and Merger Rate Distributions}
Table~\ref{tab:main} summarizes the relevant information for the {\em accelerated} runs we performed: the target event type
(NS-NS or BH-BH), the number of runs (i.e., the number of parameter combinations used), the
number of runs that were ``completed'' (i.e. which found the desired number of events of the
target type),\footnote{Almost all ``incomplete'' runs failed to find the maximum number of merging
  binaries not because the model parameters would not produce them, but rather  because 
  practical matters (e.g. computer problems)  prevented the run from finishing according to our
  desired conditions.} the number of target events $n$ needed for a run to be considered ``complete'',
and the maximum number of primordial binaries allowed in these runs.
Our sample consists of {\em an order of magnitude more points} than have been sampled before, distributed randomly throughout many  plausible combinations of the seven model parameters varied. Each model is appropriately consistent with observations of the Milky Way disk.

\begin{figure}
\resizebox{8.8cm}{!}{\includegraphics{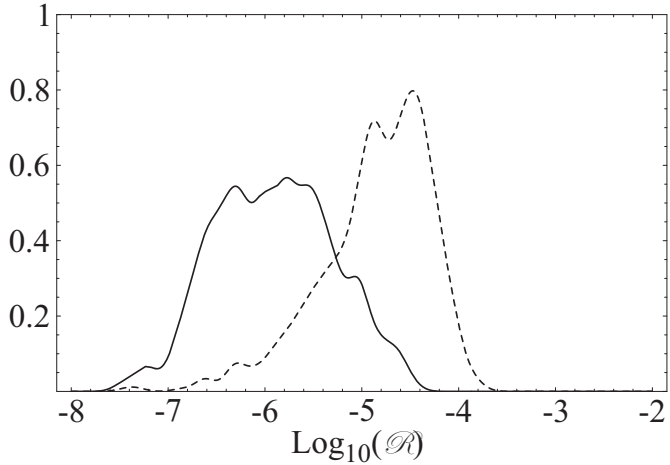}}
\caption{A smoothed, normalized histogram of $\log_{10}({\cal R})$, for ${\cal R}$ the merger
  rate for BH-BH (solid) and NS-NS (dashed)
  binaries, per year per galaxy similar to the Milky Way.  The histogram has been smoothed over a length of 0.1 in the log,
  and normalized so $\int \rho({\cal R}) d\log_{10}{\cal R} =1$.
  Our simulations are adequate to resolve the tail of each of these distributions.  For example,
  for the BH-BH simulations (cf. Table~1), our simulations have adequate
  resolution to discover models with ${\cal R}>10^{-8}$\,yr$^{-1}$.  
}
\label{fig:Histogram}
\end{figure}

\emph{Merger rate distributions}:
%
%
Figure~\ref{fig:Histogram} is a histogram of our results for the BH-BH and NS-NS merger rates
from the complete sets of runs. In practice the two histograms represent a pair of
\emph{probability distributions} of BH-BH and NS-NS inspiral rates, assuming flat prior
probability distributions for the population synthesis model parameters.

We can trust that these histograms accurately represent probability distributions because we can accurately  \emph{fit} both rate functions
over the 7-dimensional space of model parameters using the same data set.






\begin{deluxetable}{crrrrr}

\tablecolumns{6} \tablecaption{Partition table} \tablehead{ \colhead{Target Type} &
\colhead{c1} &\colhead{c2} & \colhead{c3} & \colhead{c4} & \colhead{c5}} \startdata
{\rm BH-BH} & 0.971 & 0.246 & -0.0167 & 0.1567 & 1.567 \\ 
{\rm NS-NS(b)} 
&+0.915 &+0.938 &-0.0915 &+0.1705 &+1.436 \\ 
&+0.906 &+0.689 &-0.0415 &+0.1323 &+1.323 \\ 
&-0.667 &+0.974 &-0.0293 &-0.0078 &-0.078 \\
&-0.659 &+0.018 &-0.0293 &-0.1101 &-1.101
\enddata
\label{tab:partitions}
\end{deluxetable}

\emph{\label{sec:fits}Seven-dimensional Fits for Merger Rates}:
Given the results on the NS-NS and BH-BH merger rates,
sampled randomly through the seven-dimensional  space of population synthesis model parameters,
we attempt to obtain multi-dimensional \emph{fits}. Such fits for the first time allow us 
to  rapidly estimate these two rates without
the need to resort to full population synthesis calculations that are
typically highly demanding in computing power.  The computational
  cost is dependent on the sophistication level of a given code. For
  \emph{StarTrack}, a simulation for $10^{5}$ initial binaries with
  primary masses in excess of 4\,M$\odot$ requires about 80 hours on a
  single, AMD Athlon processor of top current speed.

%
%

\begin{figure}

\resizebox{8.8cm}{!}{\includegraphics{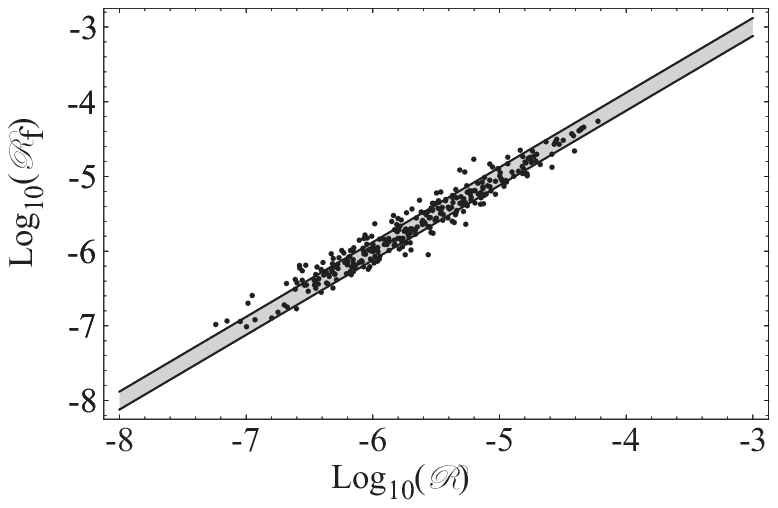}}

\resizebox{8.8cm}{!}{\includegraphics{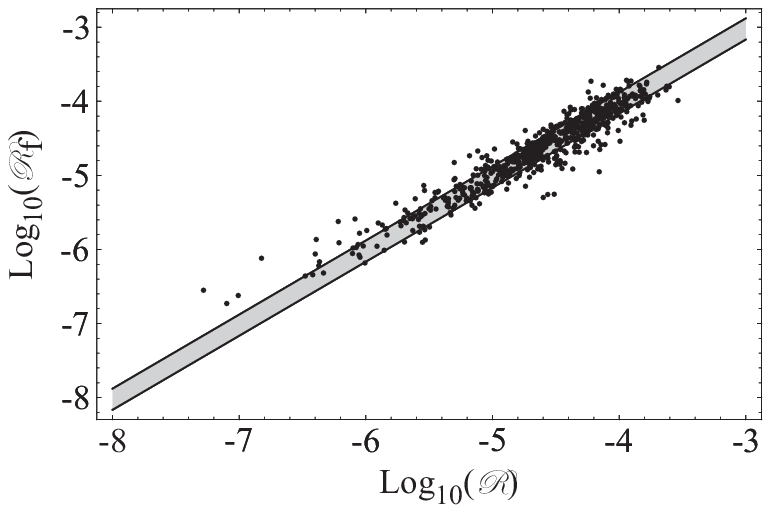}}

\caption{A plot of our fits to the BH-BH (top) and NS-NS (bottom) merger rates versus our
    original estimate of that rate.  
 Specifically, the plots show $\log_{10}{\cal
    R}_f(x_k)$ versus $\log_{10} {\cal R}_k$, where ${\cal R}_k$ is the
    numerical value of the rate at $x_k$ and $\log {\cal R}_f$ is our fit to that rate data.
   In both cases, ${\cal R}_f(x)$ was made from a quadratic fit  to the nearest $60$ points.
   The shaded region denotes our estimate for the relative uncertainty in each data point
    [i.e. the shaded region is the area between the curves ${\cal R}(1\pm 1/\sqrt{10}$].
}
\label{fig:RateErrorBH}

\end{figure}

We have used several different fitting techniques (e.g., global
polynomial fits at second and third order; non-parameteric fits using local quadratic
approximations over the nearest 60 points; \ldots).  All give comparable results: based on their residuals, the derived fits turn out to be accurate to 50\% (for NS-NS binaries) and 40\% (for
BH-BH binaries) [cf.~Figure~\ref{fig:RateErrorBH}].    In both cases, we find fits that have errors very close to the underlying errors in the data (due to statistical fluctuations  in the small number $n=10$ of merger events we observe),\footnote{The second NS-NS sample, for which $n=100$, does not have enough points by
  itself to independently produce an accurate rate estimate.  We used
  these points to test the fit obtained using the low-quality results,
  and found the errorrs were small.  We also added these
  higher-accuracy points to our lower-accuracy data to marginally
  improve our overall fit.  We found no significant change when we
  added these points.} namely $\approx 1/\sqrt{n} \approx
30\%$.


\section{Discussion}
%
%
In this paper we have systematically explored the dependence of predicted BH-BH and NS-NS
merger rates on population synthesis model parameters by (1) estimating the rate for a number of 
different combinations of model parameters that exceeds earlier parameter studies by more than an order of magnitude; and then (2) fitting to the resulting data set,
which provides us with a fast and moderately accurate ($\sim O(40-50\%)$) estimate for these
merger rates for general parameter combinations; the achieved accuracy is comparable to the underlying statistical accuracy of the runs used for the fit (30\%).

%
%

Separate from the focus of the present study, it is interesting to
point out that our extensive database of synthesis models allows us to
derive a prior probability distribution for the BH-BH merger rate,
which has not been constrained empirically.  In 
particular, this distribution clearly implies that the BH-BH merger rate is higher 
${\cal R}>10^{-8}$\,yr$^{-1}$ per galaxy similar to the Milky
Way. Such a lower limit to the rate would lead to event rates for
advanced LIGO sensitivity of at least tens of events per year. 

Furthermore, the derived fits lead to new inspiral-rate map that can immensely extend the
exploration of the relevant parameter space. We can use them to address new questions that
require a thorough understanding of how the inspiral rates vary with
parameters, such as the following:
\begin{enumerate}
\item \emph{Constraints on inputs (parameters)}: We have used only weak constraints on model
parameters: we have assumed all values in some interval to be equally likely.  Experimental
results -- most notably for the supernovae kick distribution -- can provide probability
distributions characterizing the likelihood of each parameter.  Combining these distributions
with the inspiral rate, we can better estimate the NS-NS and BH-BH rate probability distributions.

\item \emph{Constraints on outputs (rates)}: We can use the empirically derived probability
distribution for the NS-NS inspiral rate \cite{ChungleeRate1} and map it \emph{back} to the
population synthesis results. Such mapping will allow us to construct probability distributions
on the space of model parameters consistent with this empirical NS-NS distribution. We can further extend this concept and  apply many additional observational constraints simultaneously (e.g., on
certain types of supernova rates, or on the formation rate of other
known binary types, the rates of which can be empirically constrained;
cf. Kim et al. 2004) to better constrain
population synthesis model parameters and results.

\end{enumerate}

We intend to undertake the above studies as a follow-up to the development of the methods presented here. 
We expect each additional constraint will substantially reduce the set of population synthesis models consistent with observed astronomical populations.    For example, as we will present in a subsequent paper, by merely comparing a single prediction of population synthesis calculations (i.e., the NS-NS merger rate) with the known populations, we can exclude of order half our model space: half of our \emph{a priori} model parameters are inconsistent with the observed distribution of binary pulsars.   Since population synthesis calculations implicitly produce a vast number of predictions which can be compared against observed populations (e.g., the supernovae rate; the statistics of X-ray binaries; etc) we expect that a more systematic study of experimental constraints will produce very stringent constraints on physically reasonable population synthesis parameters.

%
%
Last we note that our code can and should be improved.  For example, we are considering moving away from pure
monte carlo, instead using nonrandom events (e.g., progenitor choices and supernovae kicks) and
weighted systems to better sample the relevant parameter spaces and improve convergence.  We
are also exploring more sophisticated methods to seperate the progenitors of ``irrelevant''
events from the progenitors of each target species.

\acknowledgments

This work is partially supported by a NSF Gravitational Physics grant PHYS-0121416, a David and Lucile Packard Foundation Fellowship in Science and Engineering, and a Cottrell Scholar Award
from the Research Corporation to VK.

\newpage

\appendix

\section{\label{ap:norm}A. Absolute Normalization of Synthesis Runs}

%
%
The scale factor $s$ is a ratio
$s=N_g/N_{eff}$ between the number of
stellar systems in the Milky Way and the number of stellar systems we have effectively sampled
to select our $n$ merging compact binaries.

%
%
\emph{Effective sample size}: The effective sample size $N_{eff}$ is the number of stellar
systems needed, on average, to produce $N$
stellar systems with $m_1>4$, if all systems are drawn from an IMF which extends from the
hydrogen burning limit $m=0.08 M_\odot$ to $m=150 _\odot$:
\begin{equation}
N_{eff} = N/\int_4^{150} dm \; \phi(m) \;.
\end{equation}
We use a Kroupa IMF: $\phi(m) \propto m^{-1.3}$ if $ m\in[0.08, 0.5] M_\odot$, $ \propto
m^{-2.2}$ if $ m\in[0.5, 1] M_\odot$, and $ \propto m^{-2.7}$ if $ m > 1 M_\odot $.

%
%
\emph{Estimating the number of stellar systems in the galaxy}: We choose $N_g$ so that $N_g$
times the average mass (according to our IMFs for $m_1$ and $q$ and the binary fraction $f_b$)
of each stellar system $\left< m_{tot} \right>$ is equal to the total mass which should be
formed in stars over the $T=10$Gyr lifetime of the Milky Way, $\dot{M} T$:
\begin{eqnarray}
N_g &=& \frac{\dot{M} T}{\left< m_{tot} \right>} = \frac{\dot{M} T}{ \left< m_1 \right>
     \left(1+ f_b \left<q\right>\right)}\;.
\end{eqnarray}
For the rate $\dot{M}$ at which mass is born in stars, we use the empirical estimate $\dot{M}
 \approx 3.5 M_\odot/yr$ \citep{SFR,Blitz,Lacey}.  The average values of $m_1$ and $q$ are
 found from the Kroupa IMF and Eq.~(\ref{eq:rhoq}), respectively.

\end{document}